\documentclass[conference]{IEEEtran}
\usepackage[letterpaper, top=0.7in, bottom=1.05in, left=0.631in, right=0.631in]{geometry}
\usepackage{array}
\usepackage{pgfplotstable}
\usepackage{booktabs}
\usepackage{bbm}      
\usepackage{subfig}
\usepackage{graphicx,dblfloatfix}
\usepackage{caption}
\usepackage{color}
\usepackage{amssymb}   
\usepackage{amsmath}
\usepackage{amsthm}
\usepackage{algorithm}
\usepackage{algpseudocode}
\usepackage{enumerate}
\usepackage{tabularx}  
\usepackage{multirow}
\usepackage[draft]{hyperref}      
\usepackage{url} 
\usepackage[T1]{fontenc}
\usepackage{enumerate}
\usepackage{float}
\usepackage{cite}
\usepackage{verbatim}
\usepackage{listings}
\usepackage{setspace} 
 
\begin{document}

\title{FaiRTT: An Empirical Approach for Enhanced RTT Fairness and Bottleneck Throughput in BBR}
\author{
\IEEEauthorblockN{Akshita Abrol\IEEEauthorrefmark{1}, Purnima Murali Mohan\IEEEauthorrefmark{1}, Tram Truong-Huu\IEEEauthorrefmark{1}\IEEEauthorrefmark{2}}
\IEEEauthorblockA{
\IEEEauthorrefmark{1}Singapore Institute of Technology (SIT), Singapore \\  
\IEEEauthorrefmark{2}Agency for Science, Technology and Research (A*STAR), Singapore
}
Email: \{akshita.abrol, purnima.mohan, truonghuu.tram\}@singaporetech.edu.sg
}

\maketitle  

\begin{abstract}
In next-generation networks, achieving Round-trip Time (RTT) fairness is essential for ensuring fair bandwidth distribution among diverse flow types, enhancing overall network utilization. The TCP congestion control algorithm --- BBR, was proposed by Google to dynamically adjust sending rates in response to changing network conditions. While BBRv2 was implemented to overcome the unfairness limitation of BBRv1, it still faces intra-protocol fairness challenges in balancing the demands of high-bandwidth, long-RTT elephant flows and more frequent short-RTT mice flows. These issues lead to throughput imbalances and queue buildup, resulting in elephant flow dominance and mice flow starvation. In this paper, we first investigate the limitations of Google's BBR algorithm, specifically in the context of intra-protocol RTT fairness in beyond 5G (B5G) networks. While existing works address this limitation by adjusting the pacing rate, it eventually leads to low throughput. We hence develop the FaiRTT algorithm to resolve the problem by dynamically estimating the Bandwidth Delay Product (BDP) sending rate based on RTT measurements, focusing on equitable bandwidth allocation. By modeling the $Inflight$ dependency on the BDP, bottleneck bandwidth, and packet departure time after every \texttt{ACK}, we can resolve the intra-protocol fairness while not compromising the throughput on the bottleneck link. Through extensive simulations on NS-3 and comprehensive performance evaluations, FaiRTT is shown to significantly improve the fairness index and network throughput, significantly outperforming BBRv2, for diverse flow types. FaiRTT achieves an average throughput ratio of $1.08$ between elephant and mice flows, an average fairness index of $0.98$, and an average utilization of the bottleneck link of $98.78\%$.


\end{abstract}

\begin{IEEEkeywords}
BBR, Traffic Congestion Control, Flow Fairness, Network Throughput, Elephant and Mice Flows.
\end{IEEEkeywords}

\section{Introduction}

The Bottleneck Bandwidth and Round-trip propagation time (BBR) algorithm, introduced by Google in 2016~\cite{cardwell2016bbr}, stands out as a pivotal mechanism, especially considering its capability to dynamically adjust sending rates in response to real-time network conditions. The versatility of BBR makes it an indispensable tool, not merely for enhancing throughput and reducing latency in the ever-evolving and complex network landscapes but also for effectively navigating the challenges brought forth by the continuous advancement in network technologies and their accompanying protocols. In next-generation networks (NGN) such as beyond 5G (B5G) or 6G, the analysis of flow size statistics showcases a consistent pattern of \textit{heavy-tail} behavior for applications ranging from bandwidth-demanding long flows (e.g., video-surveillance drone applications, AR/VR)  to the delay-sensitive short flows~\cite{chaccour2022seven}. Such heavy-tail flow distribution in NGN is similar to that of the Internet's familiar \textit{mice-elephant} phenomenon. 

The high-bandwidth, high Round-Trip Time (RTT) applications, which account for a smaller fraction of the overall network traffic, are referred to as the ``elephant'' flows. This contrasts sharply with the plethora of smaller ``mice'' flows, such as tactile internet signaling traffic, HTTP requests, etc., which, despite their higher frequency, low-RTT constitute a lesser segment of the total traffic volume. In the B5G network landscape, the distinct roles of high-bandwidth, long-RTT elephant flows, and frequent short-RTT mice flows are even more pronounced with their varying QoS and RTT requirements. To model the network flows, BBR works on Kleinrock’s optimal operating point~\cite{kleinrock1979power} by estimating the Maximum Delivery Rate referred to as the Bottleneck Bandwidth (\texttt{BtlBw}) and the Minimum Round-Trip Time known as \texttt{RTprop}. Based on the Bandwidth Delay Product (BDP), measured as the product of \texttt{BtlBw} and \texttt{RTprop}, the algorithm calculates three control parameters: (i) sending quantum, (ii) pacing rate, and (iii) congestion window. BBR estimates bottleneck bandwidth to set sending rates --- aiming for high throughput without increasing the waiting time of the packets in the queue~\cite{cardwell2016bbrietf}. 

However, the first version of BBR (BBRv1) overestimates the sending rate by constantly filling the BDP leading to queue buildup, bandwidth, and RTT unfairness issues~\cite{scholz2018tcpbbr,hock2017bbrevaluation,scherrer2022bbr}. BBRv2 was developed to overcome these limitations by introducing mechanisms to better estimate and adapt to the actual available bandwidth and minimize queue buildup. While BBRv2 adapts better to network changes than BBRv1, it still struggles in NGN networks with highly varying QoS requirements. Ensuring complete fairness among multiple BBRv2 flows is challenging, especially in NGN scenarios of the coexistence of elephant and mice flows due to the overestimation (elephant flows) or underestimation (mice flows) of the optimal operating point. For example, the elephant flows tend to dominate, causing unfairness and queue build-up at the router while the mice flows face starvation due to the unintended shift in the operating point due to the overestimation. Consequently, this unintended shift in the operating point leads to slow, unstable convergence, RTT unfairness, and throughput underutilization by different flow RTTs~\cite{njogu2023bbr_efra,pan2022bbrv2_improvement}. Some efforts have been made to improve the intra-protocol fairness for BBRv1 in~\cite{njogu2023bbr_efra} and BBRv2 in~\cite{pan2021acw_bbr}. They utilize queue size estimation to adjust the pacing gain to improve fairness. However, there is no emphasis on changing the BDP itself depending on the incoming flow.

In this paper, we develop a novel algorithm (named FaiRTT) that can dynamically estimate the BDP based on flow-centric RTT estimate to mitigate the observed fairness challenges. The primary objective is to ensure equal opportunity for both elephant and mice flows within the bottleneck link, addressing the dominance and starvation issues, respectively, prevalent in existing BBR implementations. We use the standardized Jain's fairness index as one of the key performance metrics to demonstrate the effectiveness of FaiRTT, combined with other network performance metrics such as throughput, and link utilization. Through the performance evaluation on NS-3, we show that FaiRTT has a better average throughput ratio of 1.08 between elephant and mice flows and an average fairness index of up to 0.98 compared to that of BBRv2 while improving the overall bottleneck link bandwidth utilization to $98.78\%$.

The rest of the paper is organized as follows. In Section~\ref{sec:related works}, we discuss the related work. We present the FaiRTT algorithm for improving RTT fairness in Section~\ref{sec:algorithm}. Section~\ref{sec:evaluation} presents our experiments and analysis of results before we conclude the paper in Section~\ref{sec:conclusion}.  

\section{Related Work}
\label{sec:related works}

Since the introduction of Google's BBR congestion control algorithm, numerous experiments and analyses have been performed to evaluate its performance in various network scenarios. In this section, we review several works and discuss potential issues that can happen to traffic flows.

\subsection{BBR Operational Intricacies}

In contrast to the initial promising results reported by Google on the effectiveness of BBRv1 (BBR-version 1), subsequent studies uncovered several issues in its behavior. Scholz \textit{et al.}~\cite{scholz2018tcpbbr} identified the creation of long-standing queues during the startup phase, suppressing existing flows. Hock \textit{et al.}~\cite{hock2017bbrevaluation} observed bandwidth discrepancies between elephant and mice flows sharing a bottleneck link due to the higher BDP of elephant flows. Scherrer \textit{et al.}~\cite{scherrer2022bbr} found that in the presence of multiple flows managed by BBRv1, overestimated delivery rates lead to standing queues exceeding 1.5 times BDP, causing packet loss and unfair bandwidth shared among flows, particularly for small buffer sizes. To address the above concerns, Google introduced BBRv2 (BBR-version 2)~\cite{cardwell2018bbrv2} in 2018, incorporating explicit congestion notification (ECN) and packet loss rate. Zhang \textit{et al.}~\cite{zhang2019bbrvariants} while evaluating various BBR versions, highlighted BBRv2's improvements in flow fairness and enhanced coexistence with CUBIC and Reno. However, suboptimal channel utilization was noted with a 5\% loss rate. Song \textit{et al.}~\cite{song2020bbrv2intraprotocol} reported fairness enhancements and reduced re-transmissions with BBRv2, emphasizing its effectiveness in limited buffer scenarios but noting challenges in fair bandwidth convergence with two flows entering a bottleneck link at different times. Nandagiri \textit{et al.}~\cite{nandagiri2020bbrv1v2} conducted an experimental evaluation comparing BBRv1 and BBRv2, revealing BBRv2's ability to overcome inflight capacity limitations in networks with small buffers. However, in networks with large buffers, long RTT flows still consume more bandwidth.

 \subsection{BBR Enhancing Fairness}
 
Ma \textit{et al.}~\cite{ma2017bbq} introduced the BBQ algorithm to enhance RTT fairness by minimizing probing periods for flows with extended round-trip times. While effective, BBQ's performance degrades with an increasing number of competing flows.
 Kim \textit{et al.} proposed solutions in~\cite{kim2019queue_bbr} and~\cite{kim2019enhancedbbr} to limit BBRv1's inflight capacity during the ProbeBW phase, addressing bias towards long RTT flows and preventing excessive data transmission. However, it causes a decrement in total throughput due to limiting inflight capacity to 1 BDP to prevent congestion. In~\cite{kim2019da_bbr}, the authors presented the Delay-Aware BBR (DA-BBR) algorithm to improve RTT fairness and throughput using the relation of each flow's RTT with \texttt{RTprop}. However, this algorithm was noted for slow and unstable convergence. 
 Njogu \textit{et al.}~\cite{njogu2023bbr_efra} introduced BBR-With Enhanced Fairness (BBR-EFRA), dynamically adjusting congestion window (CWND) based on buffer queue status for equitable competition among different RTT flows. 
Pan \textit{et al.}~\cite{pan2022bbrv2_improvement} addressed the fairness and excessive re-transmissions in BBRv2 with the introduction of BBRv2+, utilizing flow-aware explicit congestion notification. However, these algorithms rely on accurate real-time queue estimation, which is challenging in dynamic network conditions. 

Addressing these issues remains crucial for achieving equitable bandwidth allocation among different RTT flows without compromising on the throughput through a stable convergence. With the continuous updating of the BBR algorithm, BBRv2 can solve some fairness problems and limitations of BBRv1. BBRv3 (BBR-version 3)~\cite{cardwell2023bbrv3} has recently been released, aiming to rectify bugs and optimize performance parameters observed in BBRv2. While BBRv2 and BBRv3 present updates and improvements, experiments reveal persistent RTT unfairness.  Thus, continued research is essential to enhance RTT fairness and intra-protocol fairness, providing potential refinements for the final version of BBR.

\section{FaiRTT Algorithm}
\label{sec:algorithm}

\subsection{Overview of BBRv2}


Given a network, the main objective of BBR is to maximize the utilization of the bottleneck link with a minimal delay. BBR estimates the bandwidth of the bottleneck link as the maximum observed delivery rate and the propagation delay as the minimum observed RTT over periodic intervals. The bandwidth of the bottleneck link is denoted as $\texttt{BtlBw}$, and the propagation delay is denoted as $\texttt{RTprop}$. However, sending more traffic (probing more bandwidth) to estimate $\texttt{BtlBw}$ may cause congestion at the bottleneck link, thus increasing $\texttt{RTprop}$ and vice-versa. BBR, therefore, estimates these values separately in its different execution phases. 

\begin{figure}[t]
   \centering
   \includegraphics[width=0.48\textwidth]
{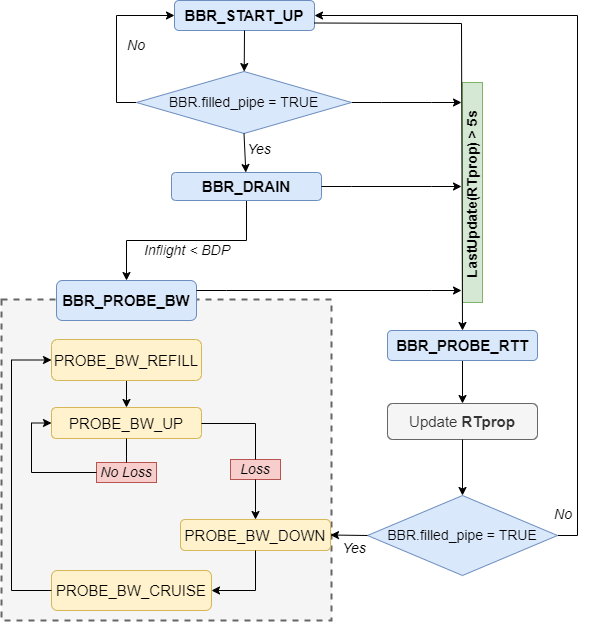}
   \caption{BBRv2 Phases and Flow Diagram.}
   \label{fig:bbrv2_flow}
\end{figure}

\begin{table}[t]
\caption{Mathematical Notations}
 \label{table:symbols}
 \footnotesize
 \begin{center}
  \begin{tabular}{|l|l|}
   \hline
   {\bf Notation} & {\bf Definition} \\
   \hline
    $BDP_t$ & Bandwidth delay product at time window $t$\\\hline
    $\texttt{RTprop}_t$ & Minimum RTT captured at time window $t$ \\\hline
    $\texttt{BtlBw}_t$ & Max. bottleneck bandwidth at time window $t$ \\\hline
    {$lastRTT_t$} & RTT of the last \texttt{ACK} at time window $t$\\\hline
     ${minRTT}_t$ & Minimum value of RTT at time window $t$\\\hline
     \multirow{2}{*}{${dRate}_t$} & Delivery rate of the bottleneck link \\ 
     & at time window $t$\\\hline 
     $maxBw_t$ & Maximum delivery rate at time window $t$ \\\hline
     \multirow{2}{*}{$Inflight_t$} & Estimated volume of in-flight data to utilize \\ 
     & available bottleneck bandwidth at time window $t$\\\hline
    $\alpha_t$ & RTT fairness threshold at time window $t$ \\\hline
    $Wfcount_t$ & Number of flows at time window $t$ \\\hline
    $WminRTT_t$ & RTT estimate at time window $t$\\\hline
    $\beta$ & Balance factor set to 0.8 \\\hline
    $\gamma$ & Discount factor set to 0.99 \\\hline
    \texttt{PROBE\_INT\_EXP} & Expiration flag of the probe bandwidth phase \\\hline
    $\texttt{RATE\_LIM\_APP}$ & Rate limiting flag \\
   \hline
\end{tabular}
\end{center}
\vspace{-2.5ex}
\end{table}

In Fig.~\ref{fig:bbrv2_flow}, we present the flow diagram of BBRv2. We provide the meaning of mathematical notations used in the BBRv2 algorithm in Table~\ref{table:symbols}. The BBRv2 algorithm is composed of four
phases: Startup (\texttt{BBR\_START\_UP}), Drain (\texttt{BBR\_DRAIN}), Probe Bandwidth (\texttt{BBR\_PROBE\_BW}), and Probe RTT (\texttt{BBR\_PROBE\_RTT}).
The first phase (\texttt{BBR\_START\_UP}) adapts the exponential Startup behavior
from CUBIC by doubling the sending rate with each RTT. The \texttt{pacing\_gain} is set to $2.89$. If the increase in the sending rate does not exceed 25\% for three consecutive attempts or if packet loss or an ECN-marked rate surpasses a predefined threshold (e.g., 2\%), BBRv2 assumes to have reached the bottleneck bandwidth. Since this observation is delayed by one RTT, a queue was already created at the bottleneck. In the second phase (\texttt{BBR\_DRAIN}), BBRv2 tries to drain the queue by
temporarily reducing the \texttt{pacing\_gain} to approximately $0.34$. Afterwards, BBRv2 enters
the third phase (\texttt{BBR\_PROBE\_BW}), in which it probes for more available bandwidth. BBRv2 specifies exit conditions for each stage of the third phase (ProbeBW Refill, ProbeBW Up, ProbeBW Down, and ProbeBW Cruise). BBRv2 initiates bandwidth probing by linearly augmenting the amount of inflight data over one $RTprop$ in the ProbeBW Refill stage, effectively filling the network pipe. Subsequently, BBRv2 engages in the ProbeBW Up phase, during which it increases the inflight data swiftly to probe for more bandwidth. This increase of inflight data ceases when the volume reaches 1.25 times the BDP, or when packet loss or the ECN rate surpasses a predetermined threshold. BBRv2 establishes \texttt{inflight\_hi} when confronted with a higher loss/ECN rate than the threshold, ensuring that it does not surpass the operational point where excessive packet loss may occur. Upon concluding the ProbeBW Up phase, BBRv2 transitions to the ProbeBW Down phase to offset the queue. It exits this phase once the volume of inflight data becomes less than BDP and maintains a constant delivery rate in ProbeBW Cruise until the initiation of the next bandwidth probing cycle. If BBRv2 experiences packet loss in the ProbeBW Cruise, it uses \texttt{inflight\_lo} to cope with the temporary packet loss. BBRv2 continuously samples the bandwidth for BDP estimation using a time window basis. At the time window $t$,  $\texttt{BtlBw}_t$ is defined as follows:
\begin{equation}
\begin{aligned}
    \texttt{BtlBw}_t &= 
    \begin{cases}
        maxBw_t, & dRate_t \geqslant \texttt{BtlBw}_{t-1}, \\
        maxBw_t, &  \texttt{RATE\_LIM\_APP} = \texttt{FALSE}, \\
        \texttt{BtlBw}_{t-1}, &  \text{otherwise}
    \end{cases}
    \end{aligned}
\label{eq:btlbw2}
\end{equation}
where $maxBw_t$ is the maximum delivery rate captured from the socket within time window $t$, normally set to two times the probe bandwidth phase. It is to be noted that $\texttt{BtlBw}_t$ is measured only for the applications with no rate limit, indicated by the flag \texttt{RATE\_LIM\_APP = FALSE}. 

After not measuring a new \texttt{RTprop} value for 5 seconds, BBRv2 stops probing for bandwidth and enters the last phase (Probe RTT). During this phase, the bandwidth is reduced to half of BDP to drain any possible queue and get a real estimation of the RTT. This phase is kept for 200 ms, controlled by the flag \texttt{PROBE\_INT\_EXP}. If a new minimum value is measured, $\texttt{RTprop}_t$ is updated as follows:
\begin{equation}
\begin{aligned}
    \texttt{RTprop}_t &= 
     \begin{cases}
        {lastRTT}_t, & 0 \leqslant {lastRTT}_t \leqslant \texttt{RTprop}_{t-1}, \\
        {lastRTT}_t, &  \texttt{PROBE\_INT\_EXP} = \texttt{TRUE}, \\
        \texttt{RTprop}_{t-1}, & \text{otherwise}
    \end{cases}
\end{aligned}
\label{eq:rtprop}
\end{equation}
where $lastRTT_t$ is captured from the network whenever an \texttt{ACK} packet is received. 

The BBRv2 algorithm estimates the amount of data to be transmitted based on the current BDP. The BDP is in turn calculated by the product of the bottleneck bandwidth at time window $t$ and the estimated RTT of the flows propagating in that time window, $\texttt{RTprop}_t$. 
\begin{equation}
    BDP_t = \texttt{BtlBw}_t \times \texttt{RTprop}_t.
    \label{eq:bdp}
\end{equation}

\subsection{Drawback of BBR}

When multiple flows with diverse RTT coexist over a shared bottleneck link, BBR tends to transmit excess data, approximately $1.5$ times the intended amount~\cite{scholz2018tcpbbr}. Consequently, BBR deviates from its typical operating point, leading to a total transmission rate exceeding the available bandwidth of the bottleneck link. This results in a persistent queue on the bottleneck link which is shared among all flows --- predominantly favoring the elephant flows, causing bandwidth unfairness.

Elephant flows, with higher BDP values, transmit more packets and utilize more bottleneck buffers. Our experimental analysis reveals that the BDP value of mice flows becomes smaller than that of elephant flows. Consequently, mice flows inject fewer packets into the pipeline, reducing the delivery rate. The decrease in delivery rates amplifies the dominance of elephant flows, leading to increased throughput imbalance, higher packet loss, re-transmission rates, and elevated latency. In BBRv1, mice flows were found to be constrained by the congestion window, thereby reducing their transmission rates. BBRv2 aimed to mitigate this issue by using ECN and setting of \texttt{inflight} upper and lower bound. However, in ~\cite{song2020bbrv2intraprotocol}, the authors reported that BBRv2 still experienced significant issues with coexistence between different RTT flows particularly with large buffers. Furthermore, in~\cite{gomezgaona2020tcpbbrv3}, the authors revealed that BBRv3 continues to exhibit RTT unfairness. Malicious actors can exploit this vulnerability, deliberately increasing delay to gain a disproportionately higher bandwidth share. 

\vspace{-2ex}
\subsection{FaiRTT Algorithm}

\vspace{-2.5ex}
\begin{algorithm}[h]
\small 
\caption{FaiRTT: RTT Fairness Improvement Algorithm}
\label{alg:rtt_fairness}
\begin{algorithmic}[1]
     \State {\bfseries Input:} $\texttt{BtlBw}_t$, $\texttt{RTprop}_t$, $cwnd\_gain$, $lastRTT_t$\State {\bfseries Input:} $minRTT_t$, $\gamma$
     \State {\bfseries Output:} $Inflight_t$
     \State \(Wfcount_t \gets \texttt{length}(\texttt{unique}(minRTT) \text{ in } \text{window } t)\)
     \State Calculate $\alpha_t$ using Eq.~\eqref{eq:alpha}
     \If{\(\text{BBR} \rightarrow \text{mode} == \text{BBR\_PROBE\_BW}\)}
        \If{\(lastRTT_t > \alpha_t \)}
             \State \(BDP_t \gets \texttt{BtlBw}_t \times \texttt{RTprop}_t \times \displaystyle\left(\frac{minRTT_t\times \gamma}{lastRTT_t}\right) \)

        \Else
        \State \(BDP_t \gets \texttt{BtlBw}_t \times \texttt{RTprop}_t\)
        \EndIf
    \Else
    \State \(BDP_t \gets \texttt{BtlBw}_t \times \texttt{RTprop}_t\)
    \EndIf
    \State $Inflight_t \gets BDP_t \times cwnd\_gain$
    \State \Return $Inflight_t$
\end{algorithmic}
\end{algorithm}
\vspace{-1.5ex}

To address the inherent design issue of BBR, we develop Algorithm~\ref{alg:rtt_fairness}, which dynamically captures the changes in RTT and adaptively adjusts inflight for each BBR flow request. The proposed algorithm aims to improve RTT fairness while maintaining the overall throughput of the network. We leverage the calculated RTT from each acknowledgment packet (\texttt{lastRTT}) and minimum RTT (\texttt{minRTT}) to calculate an adjustment coefficient for curtailing the amount of inflight data for each flow. The inflight is defined as follows:
\begin{equation}
    {Inflight}_t = {BDP}_t \times cwnd\_gain
    \label{eq:inflight}
\end{equation}
where $cwnd\_gain$ is fixed at $0.5$ for the \texttt{BBR\_PROBE\_RTT} phase and at $2$ for all other phases of BBRv2. While all BBR flows exceed their intended inflight data transmission, the magnitude differs, particularly with elephant flows transmitting significantly larger volumes. To account for this, we examine the relationship between \texttt{lastRTT} and \texttt{minRTT}, introducing an adjustment coefficient with a value below $1$. This coefficient will approach $1$ for mice flows, while for elephant flows it would have a smaller value. The BDP calculation incorporates the adjustment coefficient, dynamically adjusting inflight data based on RTT. The adjustment coefficient is multiplied with the BDP only when the value of \texttt{lastRTT} is more than the RTT fairness threshold $\alpha_t$, calculated as:
\begin{equation}
\vspace{-0.5ex}
    \alpha_t = {Wfcount}_t \cdot \sum\nolimits_{j \in W_t} {WminRTT}_j/W_t
    \label{eq:alpha}
\end{equation}
where 
\begin{equation}
    {WminRTT}_t = \beta \cdot  \sum_{i={0}}^{t-1} \frac{{minRTT}_i}{t} + (1-\beta) \cdot {minRTT}_t
\end{equation}
and
\begin{equation}
     W_t = \min(|\mathbb{R}\texttt{x}|,|\mathbb{T}\texttt{x}|)
\end{equation}
where $|\mathbb{R}\texttt{x}|$ and $|\mathbb{T}\texttt{x}|$ are the size of the receiving window and sending window
, respectively. $Wfcount_t$ is the estimated number of BBR flows in the time window $t$ calculated by counting the number of unique $minRTT$'s in that time window, and $\beta$ is a constant fixed at 0.8 to balance the estimate. Further, we introduce a discount factor $\gamma$ of $0.99$ to stabilize BDP and promote equitable bandwidth-sharing. Elephant flows exhibit a lower adjustment coefficient than mice flows, resulting in a more substantial reduction in elephant flow's BDP compared to that of mice flows. Consequently, by transmitting less data than the original BBR, elephant flows experience a significant reduction in buffer occupancy, thereby enhancing RTT fairness.

\section{Performance Evaluation}
\label{sec:evaluation}

\subsection{Experimental Setup}
\begin{figure}[h]
  \centering  \includegraphics[width=0.8\columnwidth]{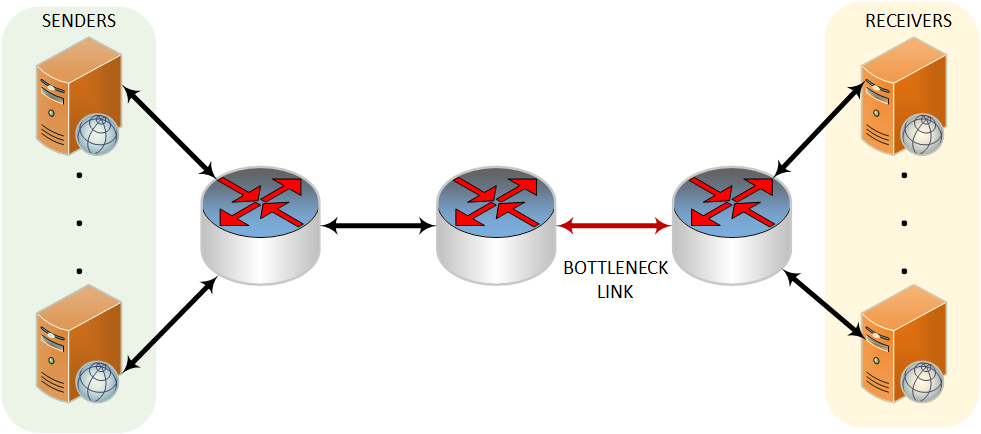}
  \caption{Experimental network topology.}
  \label{fig:topo}
  \vspace{-1ex}
\end{figure}

We used a dumbbell topology setup in the NS-3 simulation framework, featuring multiple flows sharing a common bottleneck link, as illustrated in Fig.~\ref{fig:topo}. We applied the default active queue management as the Drop-Tail policy and every packet size as 1 kB. Each simulation lasted 120 seconds and was repeated 5 times with different random seeds for the error rate in packet delivery. The obtained results were plotted with 95\% confidence interval. 

\subsubsection{Performance Metrics}

We used the following three performance metrics to evaluate the proposed algorithm.
\begin{itemize}
    \item Throughput: We used average throughput to assess the equal distribution of bandwidth among competing flows. We defined the throughput ratio as the ratio of the throughput of elephant flows and that of mice flows.
    \item Fairness Index: We used Jain's fairness index~\cite{jainindex} to provide a more comprehensive evaluation of fairness in the allocation of bandwidth among competing flows. The Jain's fairness index, $F$ is calculated using the formula:
\begin{equation}
    F = \left(\sum\nolimits_{i=1}^{n} T_i\right)^2/\left(n \cdot \sum\nolimits_{i=1}^{n} T_i^2\right)
    \label{eq:fairness_index}
\end{equation}

Here, $n$ represents the number of competing flows, and $T_i$ denotes the throughput of flow $i$. A value of $F$ close to 1 signifies fairness in the allocation of bandwidth sharing among competing flows.
    \item Bottleneck link utilization: Link utilization is vital for assessing network efficiency, measuring how effectively the bottleneck link handles data transmission.
    The percentage bottleneck link utilization $U$ is computed as:
\begin{equation}
   U (\%) = \frac{\sum_{i} \text{T}_i}{C_{\text{Btl}}} \cdot 100
    \label{eq:utilization}
\end{equation}
where $T_i$ is the throughput of flow $i$ traversing the bottleneck link and \(C_{\text{Btl}}\) is the bandwidth of the bottleneck link set to 10 Mbps for all experiments. Higher link utilization indicates better use of the network resources.

\end{itemize}
\subsubsection{Comparisons}
We compare FaiRTT with BBRv2 using above performance metrics in different experimental scenarios:
\begin{itemize}
    \item Throughput with Time
    \item Fairness Index with Time
    \item Bottleneck link utilization with Time
    \item Throughput with Queue size
    \item Throughput with RTT
\end{itemize}

\subsection{Analysis of Results}

\subsubsection{Throughput}

\begin{figure*}[t]
   \centering
   \subfloat[FaiRTT vs. BBRv2.]{\includegraphics[width=0.32\textwidth]{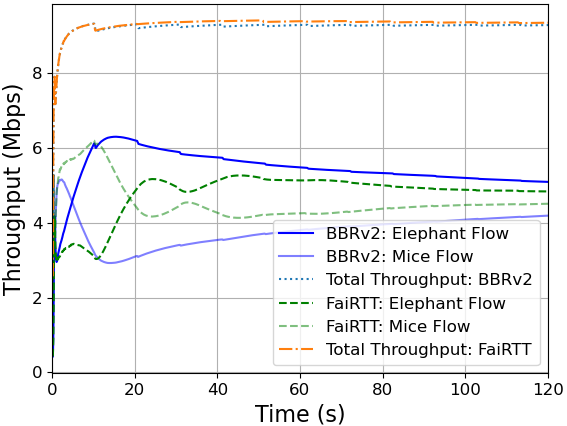}\label{fig:thput_vs_time}}
    \subfloat[With different Queue Size Values.]{\includegraphics[width=0.32\textwidth]{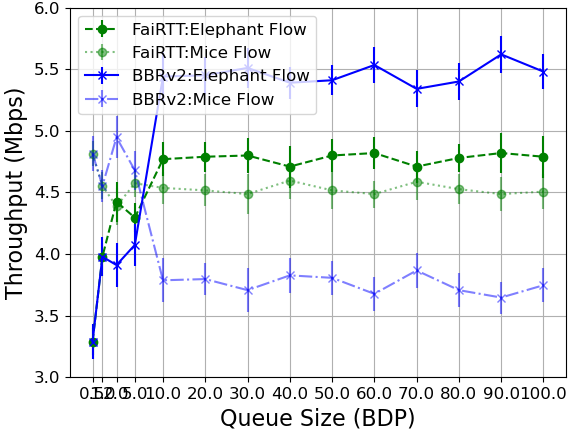}\label{fig:thput_vs_queue}}
    \subfloat[With different RTT Values.]{\includegraphics[width=0.32\textwidth]{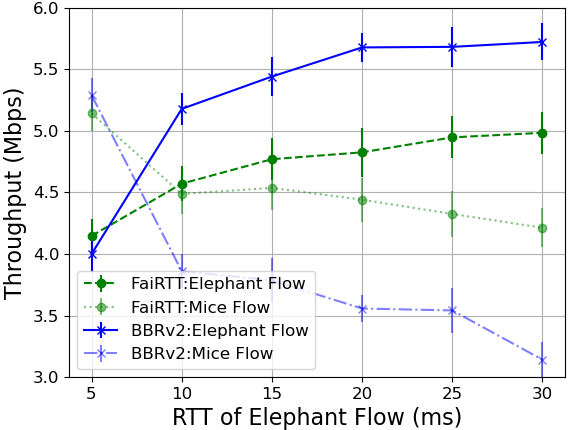}\label{fig:thput_vs_rtt}}
    \caption{Throughput Comparison.}
    \label{fig:throughput}
    \vspace{-3ex}
\end{figure*}

In this experiment, we set a bottleneck bandwidth of 10 Mbps, a delay of 10 ms, and a bottleneck buffer size of 10 BDP. The RTT of the elephant and mice flows was set to 15ms and 5ms, respectively. 
From Fig.~\ref{fig:thput_vs_time}, we can observe that regardless of the algorithms, elephant flows occupy a larger portion of throughput. BBRv2 presents highly suppressed mice flows by elephant flows with an average throughput ratio of 1.44 between the two competing flows in contrast to 1.05 in FaiRTT while maintaining the total throughput of the network. The average total throughput increased by 1\% for FaiRTT as compared to BBRv2, respectively. Hence, FaiRTT is shown to be much-needed fairness when elephant and mice flows compete for the bottleneck bandwidth. 

To evaluate the effect of bottleneck queue size on fairness, we configured the test topology having 10 Mbps bottleneck bandwidth and 10 ms bottleneck delay. The RTT of the elephant and mice flows was set to 15ms and 5ms, respectively. The bottleneck queue size was varied from 0.5 BDP to 100 BDP. Fig.~\ref{fig:thput_vs_queue} shows the throughputs of both elephant and mice flows for BBRv2 and FaiRTT with respect to the queue size. It is observed that fairness tends to decrease for larger queue sizes for BBRv2. The elephant flows occupy a much larger portion of the bottleneck bandwidth compared to mice flows particularly as the queue size increases from 10 BDP. For FaiRTT, both flows maintain almost stable and equitable throughput share for queue sizes more than 10 BDP. The algorithm performs best in terms of equitable throughput share at 2 BDP. For values less than 2 BDP, the performance of both algorithms is similar. The average throughput ratio between the competing flows is 1.46 for BBRv2 and 1.04 for FaiRTT.

We further evaluated the performance of FaiRTT for different RTTs. The RTT of elephant flows was varied from 5 to 30 ms and that of the mice flows was fixed to 5 ms. Fig.~\ref{fig:thput_vs_rtt} displays throughput changes of both algorithms when mice flows compete with elephant flows with different RTTs for a 10 Mbps bottleneck bandwidth, 10 ms delay on a 10 BDP buffer size. It can be observed that as the RTT difference between the flows increases, the fairness between the flows deteriorates. The minimum and maximum throughput ratio between the competing flows 1.02 and 1.18 for FaiRTT considerably reduces as compared to 1.32 and 1.82 for BBRv2. The overall average throughput ratio is 1.11 for FaiRTT and 1.44 for BBRv2. 

\subsubsection{Jain's Fairness Index}

\begin{figure*}[t]
   \centering
   \subfloat[FaiRTT vs. BBRv2.]{\includegraphics[width=0.32\textwidth]{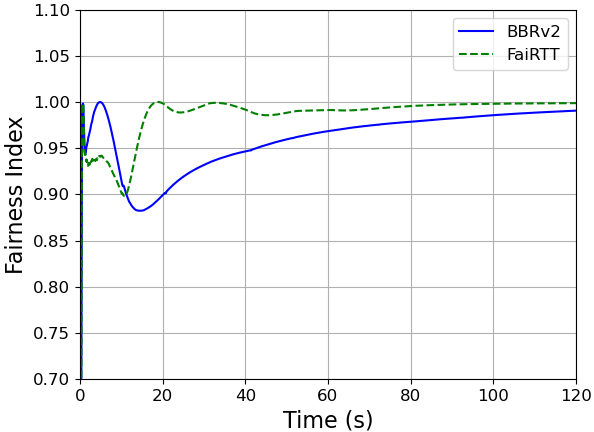}\label{fig:fairness_vs_time}}
    \subfloat[With different RTT Values.]{\includegraphics[width=0.32\textwidth]{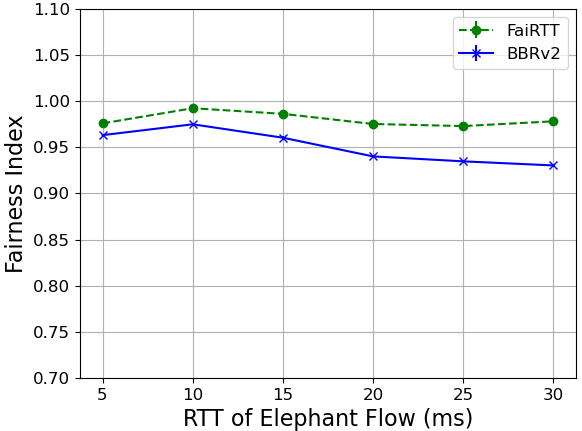}\label{fig:fairness_vs_rtt}}
    \subfloat[With different Queue Size Values.]{\includegraphics[width=0.32\textwidth]{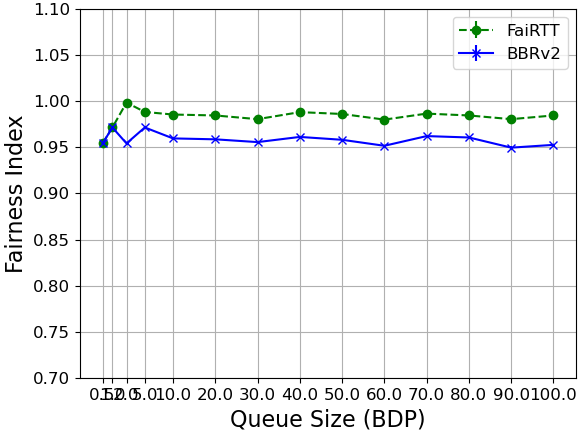}\label{fig:fairness_vs_queue}}
    \caption{Fairness Index.}
    \label{fig:fairnessindex}
    \vspace{-3ex}
\end{figure*}

Fig.~\ref{fig:fairness_vs_time} illustrates Jain's fairness index evaluation for both algorithms. We set bottleneck bandwidth at 10 Mbps and 10ms of delay. Elephant and mice flows have an RTT of 15 ms and 5 ms, respectively and they coexist on a 10 BDP bottleneck buffer size. For BBRv2, we recorded a minimum fairness index of about 0.88  and an average of 0.95 as compared to FaiRTT with 0.91 and 0.99. 
Fig.~\ref{fig:fairness_vs_queue} shows the effect of varying bottleneck queue size from 0.5 BDP to 100 BDP on the fairness index for both algorithms. The fairness index exhibits the maximum value at 2 BDP for FaiRTT at 0.99 as compared to 0.95 for BBRv2. Both algorithms exhibit the same fairness for values less than 1 BDP. The average fairness index of FaiRTT and BBRv2 are 0.99 and 0.95, respectively.
Fig.~\ref{fig:fairness_vs_rtt} illustrates the fairness index amongst competing elephant and mice flows with varying RTTs. Despite the fairness indexes’ slight decrement when competing with elephant flows with increasing RTT, FaiRTT outperforms BBRv2. For competing flows of 5 ms and 30 ms, FaiRTT has a fairness index of 0.99 versus 0.93 for BBRv2. The maximum value of the fairness index obtained for FaiRTT is 0.99 as against 0.97 for BBRv2 at 5 ms and 10 ms competing flows. The average fairness index for FaiRTT and BBRv2 are 0.98 and 0.94, respectively.
This demonstrates that FaiRTT enhances Jain’s fairness index and can guarantee improved fairness amongst elephant and mice flows in different scenarios. 

\subsubsection{Bottleneck link utilization}

\begin{figure}[t]
   \centering
   \includegraphics[width=0.32\textwidth]{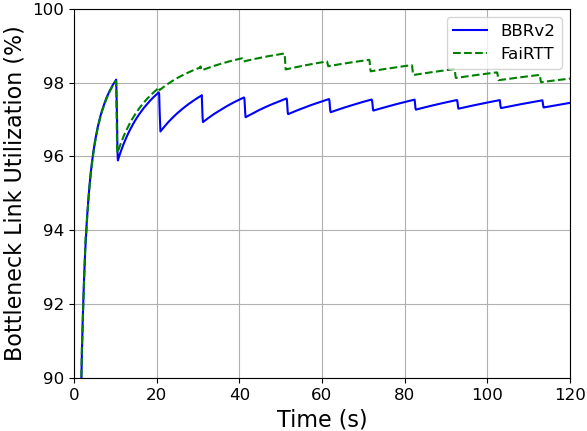} 
   \caption{Bottleneck Link Utilization of FaiRTT vs. BBRv2.}
   \label{fig:util_vs_time}
   \vspace{-3ex}
\end{figure}

Fig.~\ref{fig:util_vs_time} shows the bottleneck bandwidth utilization with the bandwidth set to 10 Mbps, delay at 10 ms, and a bottleneck buffer size of 10 BDP. FaiRTT achieves an average link utilization of 98.78\% as compared to 97.21\% in BBRv2. Hence, our proposed algorithm demonstrated its ability to optimize and utilize the available bottleneck bandwidth efficiently as compared to BBRv2. 

\section{Conclusion}
\label{sec:conclusion}

In this paper, we developed FaiRTT, a novel algorithm that dynamically estimates the BDP sending rate based on RTT measurements to achieve equitable bandwidth allocation between elephant and mice flows, addressing the drawback of BBR. In FaiRTT, we modeled the in-flight dependency on BDP, bottleneck bandwidth, and packet departure time after every \texttt{ACK}. We evaluated the performance of FaiRTT through extensive simulation experiments on multiple flows with different RTTs and buffer sizes using NS-3. The results show that FaiRTT has a better average throughput ratio of 1.08 between elephant and mice flows and an average fairness index of up to 0.98 compared to that of BBRv2 while improving the overall bottleneck link bandwidth utilization to $98.78\%$. By effectively resolving intra-protocol RTT fairness without compromising throughput on the bottleneck link, FaiRTT emerges as a promising solution to enhance overall network utilization and address the challenges posed by varying flow characteristics in next-generation networks. The findings of this study contribute valuable insights to the ongoing efforts in optimizing BBR for the evolving landscape of communication networks.

\section*{Acknowledgment and Disclaimer}
This research is supported by the Ministry of Education, Singapore, under its Academic Research Tier 1 (Grant WBS number: R-R12-A405-0003). 

Any opinions, findings, conclusions, or recommendations expressed in this material are those of the author(s) and do not reflect the views of the Ministry of Education, Singapore.

\bibliographystyle{IEEEtran}
\bibliography{bbr.bib}

\begin{thebibliography}{10}
\providecommand{\url}[1]{#1}
\csname url@samestyle\endcsname
\providecommand{\newblock}{\relax}
\providecommand{\bibinfo}[2]{#2}
\providecommand{\BIBentrySTDinterwordspacing}{\spaceskip=0pt\relax}
\providecommand{\BIBentryALTinterwordstretchfactor}{4}
\providecommand{\BIBentryALTinterwordspacing}{\spaceskip=\fontdimen2\font plus
\BIBentryALTinterwordstretchfactor\fontdimen3\font minus \fontdimen4\font\relax}
\providecommand{\BIBforeignlanguage}[2]{{%
\expandafter\ifx\csname l@#1\endcsname\relax
\typeout{** WARNING: IEEEtran.bst: No hyphenation pattern has been}%
\typeout{** loaded for the language `#1'. Using the pattern for}%
\typeout{** the default language instead.}%
\else
\language=\csname l@#1\endcsname
\fi
#2}}
\providecommand{\BIBdecl}{\relax}
\BIBdecl

\bibitem{cardwell2016bbr}
N.~Cardwell \emph{et~al.}, ``{BBR: Congestion-based Congestion Control},'' \emph{ACM Queue}, vol.~14, no.~5, 2016.

\bibitem{chaccour2022seven}
C.~Chaccour \emph{et~al.}, ``{Seven defining features of terahertz (THz) wireless systems: A fellowship of communication and sensing},'' \emph{IEEE Communications Surveys \& Tutorials}, vol.~24, no.~2, pp. 967--993, 2022.

\bibitem{kleinrock1979power}
L.~Kleinrock, ``{Power and Deterministic Rules of Thumb for Probabilistic Problems in Computer Communications},'' in \emph{IEEE ICC}, 1979.

\bibitem{cardwell2016bbrietf}
N.~Cardwell \emph{et~al.}, ``{BBR Congestion Control},'' in \emph{IETF 97th Meeting}, 2016.

\bibitem{scholz2018tcpbbr}
D.~Scholz \emph{et~al.}, ``{Toward a Deeper Understanding of {TCP BBR} Congestion Control},'' in \emph{IFIP Networking}, 2018.

\bibitem{hock2017bbrevaluation}
M.~Hock \emph{et~al.}, ``{Experimental Evaluation of {BBR} Congestion Control},'' in \emph{Proc. International Conference on Network Protocols (ICNP)}, 2017.

\bibitem{scherrer2022bbr}
S.~Scherrer \emph{et~al.}, ``{Model-based Insights on the Performance, Fairness, and Stability of {BBR}},'' in \emph{ACM IMC 2022}, 2022.

\bibitem{njogu2023bbr_efra}
C.~K. Njogu \emph{et~al.}, ``{{BBR-With Enhanced Fairness (BBR-EFRA)}: A New Enhanced {RTT} Fairness for {BBR} Congestion Control Algorithm},'' \emph{Computer Communications}, vol. 200, pp. 95--103, 2023.

\bibitem{pan2022bbrv2_improvement}
W.~Pan \emph{et~al.}, ``{Improvement of {BBRv2} Congestion Control Algorithm Based on Flow-aware {ECN}},'' \emph{Sec. and Commun. Netw.}, Jan. 2022.

\bibitem{pan2021acw_bbr}
------, ``{Improved {RTT} Fairness of {BBR} Congestion Control Algorithm Based on Adaptive Congestion Window},'' \emph{Electronics}, vol.~10, no.~5, 2021.

\bibitem{cardwell2018bbrv2}
N.~Cardwell \emph{et~al.}, ``{{BBRv2}: A Model-based Congestion Control},'' in \emph{Proc. IETF 102th Meeting}, 2018.

\bibitem{zhang2019bbrvariants}
S.~Zhang, ``An evaluation of {BBR} and its variants,'' \emph{arXiv preprint arXiv:1909.03673}, 2019.

\bibitem{song2020bbrv2intraprotocol}
Y.-J. Song \emph{et~al.}, ``{Intra-protocol Convergence Problem in {BBRv2}'s Bandwidth Probing},'' in \emph{ICTC 2020}, 2020, pp. 1016--1018.

\bibitem{nandagiri2020bbrv1v2}
A.~Nandagiri \emph{et~al.}, ``{{BBRv1 vs BBRv2}: Examining Performance Differences through Experimental Evaluation},'' in \emph{IEEE LANMAN 2020}, 2020.

\bibitem{ma2017bbq}
S.~Ma \emph{et~al.}, ``Fairness of congestion-based congestion control: Experimental evaluation and analysis,'' \emph{arXiv preprint arXiv:1706.09115}, 2017.

\bibitem{kim2019queue_bbr}
G.-H. Kim \emph{et~al.}, ``{Fairness Improvement of {BBR} Congestion Control Algorithm for Different {RTT} Flows},'' in \emph{ICEIC 2019}, 2019.

\bibitem{kim2019enhancedbbr}
------, ``{Enhanced {BBR} Congestion Control Algorithm for Improving {RTT} Fairness},'' in \emph{ICUFN 2019}, 2019, pp. 358--360.

\bibitem{kim2019da_bbr}
G.-H. Kim and Y.-Z. Cho, ``{Delay-aware {BBR} Congestion Control Algorithm for {RTT} Fairness Improvement},'' \emph{IEEE Access}, vol.~8, 2019.

\bibitem{cardwell2023bbrv3}
N.~Cardwell \emph{et~al.}, ``{{BBRv3}: Algorithm Bug Fixes and Public Internet Deployment},'' in \emph{Proc. IETF 117th Meeting}, 2023.

\bibitem{gomezgaona2020tcpbbrv3}
J.~A. Gomez~Gaona, E.~Kfoury, J.~Crichigno, and G.~Srivastava, ``Evaluating {TCP BBRv3} performance in wired broadband networks,'' 2023.

\bibitem{jainindex}
R.~Jain \emph{et~al.}, ``A quantitative measure of fairness and discrimination,'' Digital Equipment Corporation, Hudson, MA, USA, Tech. Rep., 1984.

\end{thebibliography}

\end{document}